\documentclass[cits]{PoS}
\usepackage{subfigure}
\usepackage{amssymb}
\usepackage{fontenc}
\usepackage{times}
\usepackage{mathptmx}

\newcommand{\araa}{ARA\&A}

\title{EVN observations of the Ultra Luminous Infrared Galaxies IRAS 23365+3604 and IRAS 07251-0248 }

\ShortTitle{EVN observations of the Ultra Luminous Infrared Galaxies IRAS 23365+3604 and IRAS 07251-0248 }

\author{\speaker{Cristina Romero-Canizales}
         \thanks{CRC, MAPT and AA acknowledge support from grant AYA2006-14986-C02-01, sponsored by the \textit{Ministerio de Educaci\'on y Ciencia}, Spain. This work has also benefited from research funding from the  European Community's sixth Framework Programme under RadioNet R113CT 2003 5058187.  
\emph{The European VLBI Network is a joint facility of European, Chinese, South African and other radio astronomy institutes funded by their national research councils.}}\\
         Instituto de Astrof\'{\i}sica de Andaluc\'{\i}a - CSIC, 18008 Granada, Spain\\
        E-mail: \email{cromero@iaa.es}}

\author{Miguel \'Angel P\'erez-Torres\\
        Instituto de Astrof\'{\i}sica de Andaluc\'{\i}a - CSIC, 18008 Granada, Spain \\
        E-mail: \email{torres@iaa.es}}

\author{Antxon Alberdi\\
        Instituto de Astrof\'{\i}sica de Andaluc\'{\i}a - CSIC, 18008 Granada, Spain \\
        E-mail: \email{antxon@iaa.es}}

\abstract{We present high-sensitivity, high-resolution images of the Ultraluminous Infrared Galaxies (ULIRG; L$_{\mathrm{FIR}} > 10^{12}$ L$_\odot$) IRAS 23365+3604 and IRAS 07251-0248, taken with the EVN at 6 and 18 cm. The images show a large number of compact components, whose luminosities are typical of Type IIL and Type IIn Radio Supernovae (RSNe). Further observations of these ULIRGs  will allow us to confirm, or to rule out, their nature.
The present observations are part of a project that should result in a significant number of SN detections, providing a direct measurement of the Core Collapse Superova (CCSN) rate and allowing us to estimate the Star Formation Rate (SFR) in our sample of ULIRGs (see  \cite{miguel}).}

\FullConference{The 9th European VLBI Network Symposium on The role of VLBI in the Golden Age for Radio Astronomy and EVN Users Meeting\\
		 September 23-26, 2008\\
		 Bologna, Italy}

\begin{document}

\section{Observations}

We carried out phase-reference observations of these ULIRGs using the EVN+MERLIN at 6 and 18 cm. Nine stations participated in the EVN observations at both 6 and 18 cm. Each observing epoch consisted of approximately 3.5 hours on target at each frequency, using a data recording rate of 1 Gb/sec,
thus ensuring a good u-v coverage and a low off-source noise (RMS).

J2333+3901 was the phase-reference source for the target source IRAS 23365+3604 while 3C286, 3C454.3 and 2134+004 were used as fringe finders and bandpass calibrators. We have cleaned the delay and rate solutions from the contribution of the phase-reference source structure. Significant editing, mostly at 18 cm, was also needed to minimize RFI effects. In the case of IRAS 07251-0248, J0730-0241 was the phase-reference source and 3C286 and 3C273 were used as fringe finders and bandpass calibrators.

\section{IRAS 23365+3604 \& IRAS 07251-0248}  

IRAS 23365+3604 is an advanced merger at a distance of 252 Mpc with a very high luminosity (log(L$_{\mathrm{FIR}} /$L$_\odot $) $=$ 12.13). IRAS 07251-0248 is a  disturbed and/or interacting pair of galaxies  at a distance of 344 Mpc and having a log(L$_{\mathrm{FIR}} /$L$_\odot $) $=$ 12.32. Following \cite{condon92}, the expected SN rate for those ULIRGs is approximately 5 and 8 SN yr$^{-1}$, respectively.

\section{Results and discussion}

\begin{figure}[h!]
  \begin{center}
    \subfigure[]{
         \label{fig:ir23c}
         \includegraphics[angle=0, width=4.5cm, height=3.7cm]{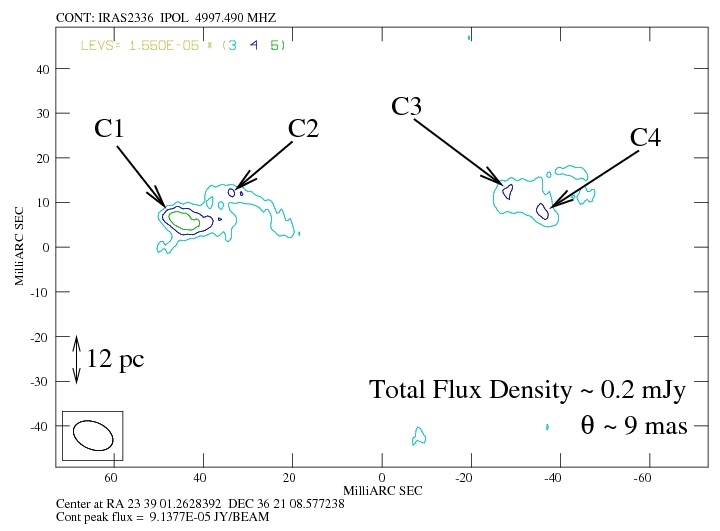}}
    \subfigure[]{
         \label{fig:ir23l}
         \includegraphics[angle=0, width=4.5cm, height=3.7cm]{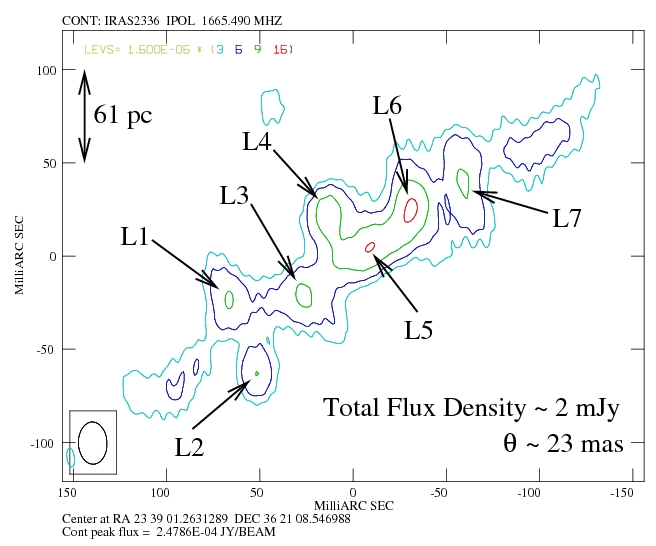}}         
    \subfigure[]{
         \label{fig:ir23cl}
         \includegraphics[angle=0, width=4.5cm, height=3.7cm]{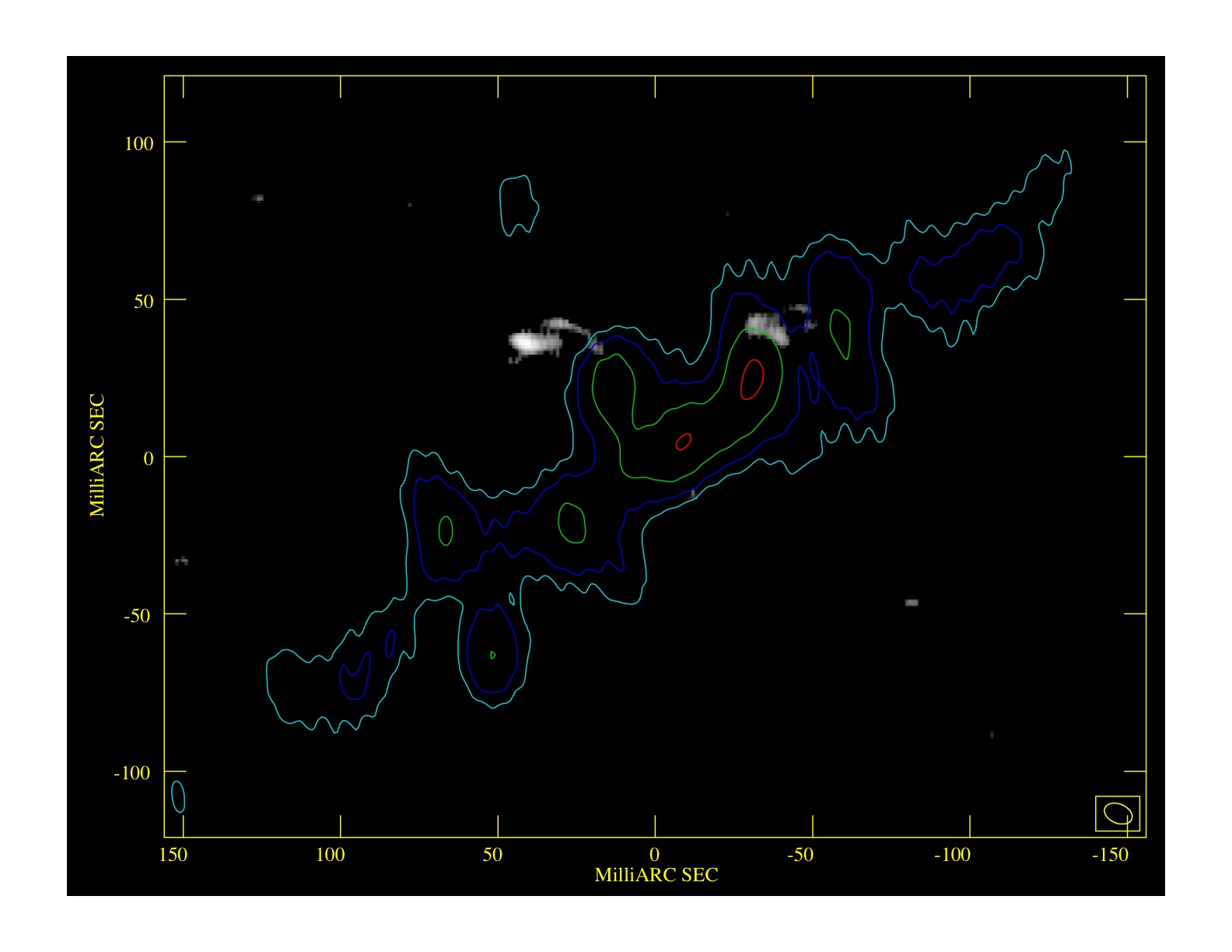}} 
            \vspace{.1in}
    \subfigure[]{                  
         \label{fig:ir07c}
         \includegraphics[angle=0, width=4.5cm, height=3.7cm]{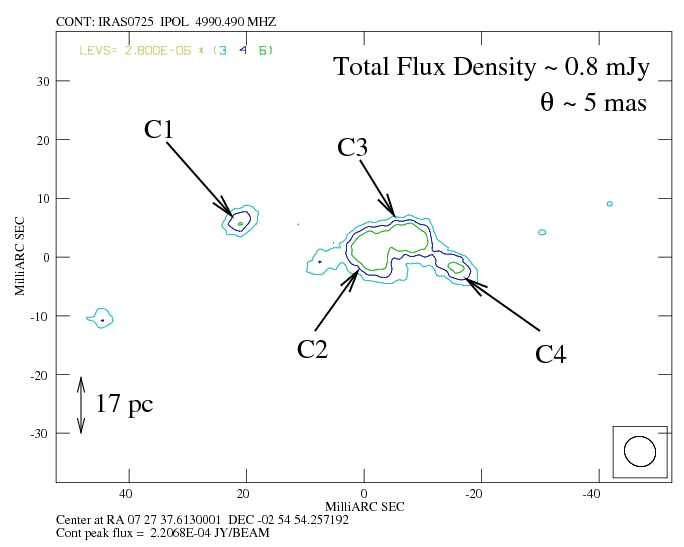}}
    \subfigure[]{
         \label{fig:ir07l}
         \includegraphics[angle=0, width=4.5cm, height=3.7cm]{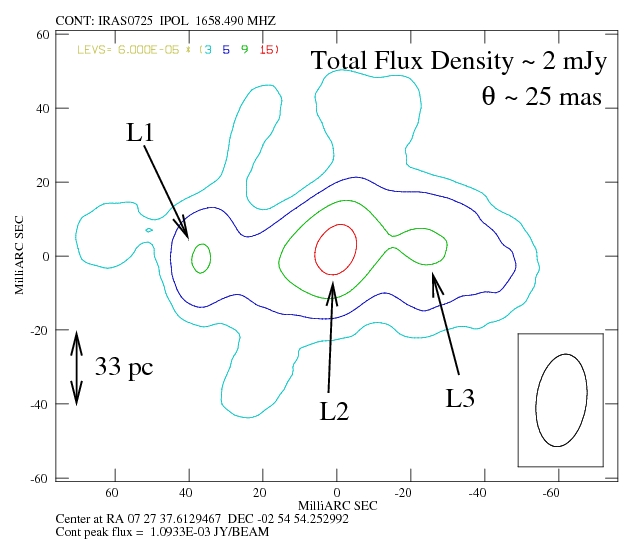}}
     \subfigure[]{
         \label{fig:ir07cl}
         \includegraphics[angle=0, width=4.5cm, height=3.7cm]{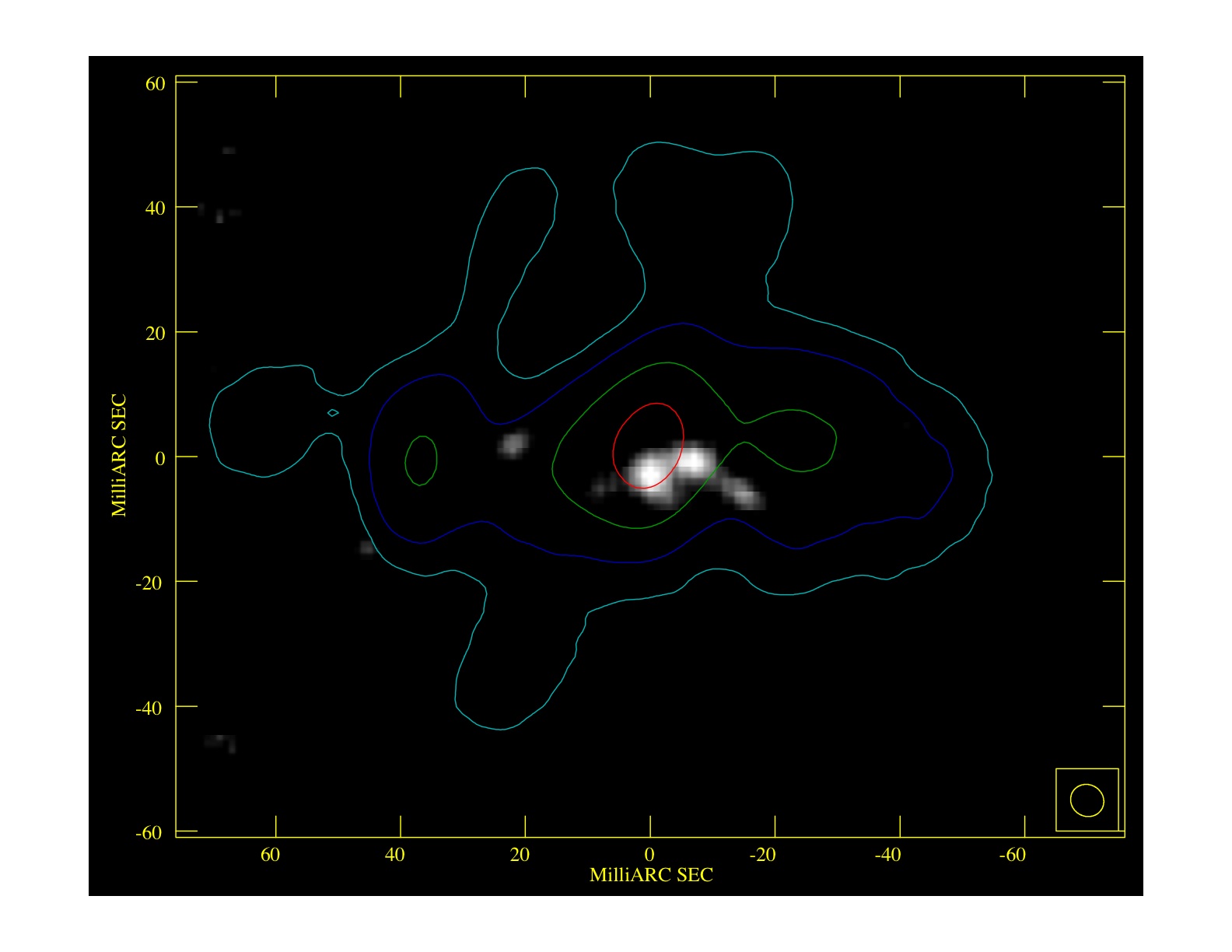}}      
   \caption { \footnotesize{IRAS 2336+3604 at 6 (a) and 18 cm (b), with estimated RMS of about 15 $\mu$Jy in both maps. IRAS 07251-0248 at 6 (d) and 18 cm (e), with estimated RMS of 28 and 60 $\mu$Jy respectively. Emission components are marked with C- and L- labels; note that the first component in the C-maps do not correspond with the first component in the L-maps and so on. IRAS 23365+3604 (c) and IRAS 07251-0241(f) contours from observations on 29 February 2008 at 18 cm overlaid on the 6 cm image from 11 March 2008 observations.} }
   \label{fig:maps}
  \end{center}
\end{figure}

We have obtained the deepest and highest resolution radio images ever of two of the most distant ULIRGs in the local Universe, using quasi-simultaneous observations with the EVN at 6 and 18 cm (see Figure \ref{fig:maps}). We expect to get a better RMS and u-v coverage in our maps once we add the MERLIN observations. 

We found a number of bright, compact components, some of which are suggestive of CCSNe exploding in the innermost regions of those ULIRGs. Indeed, the luminosities inferred for the compact components seen at 6 cm range from $\sim 5 \times 10^{27}$ in IRAS 23365+3604 up to $\sim 16 \times 10^{28}$ erg s$^{-1}$ Hz$^{-1}$ in IRAS 07251-0248 which correspond to luminosities typical of Type IIL and Type IIn SN (bright radio events). We also show overlays of the 6 and 18 cm EVN observations in Figures \ref{fig:ir23cl} and \ref{fig:ir07cl}. Not all of the 6 cm brightness peaks are coincident with those seen at 18 cm (see Table \ref{ta:param}). This is consistent with an scenario where we observe very young SNe, so that their 6 cm emission would now be
around their peak, while their 18 cm emission could still be
rising. On the other hand, we have also found several 18 cm peaks
without a clear counterpart at 6 cm. This can be explained if their
emission arises from CCSNe that are already in their optically thin
phase, as indicated by their two-point spectral indices.

Our results show that the EVN is a powerful tool to study the central regions of nearby ULIRGs, and can be efficiently used to determine the CCSN rate and SFR in those galaxies. A new observing epoch should allow us to disentangle the nature of the compact objects we have found.

\begin{table}
\begin{center}
{\scriptsize
\begin{tabular}{|c|c|c|c|c|c|c|c|c|}
\hline                        
Component & \multicolumn{4}{|c|}{IRAS23365+3604}   & \multicolumn{4}{c|}{IRAS07251-0248}                                                                                                           \\
\cline{2-9}
~  & \multicolumn{2}{c|}{Coordinates (J2000)} & $\sigma_{\mathrm{position}}$  & L$_{\nu} \times 10^{28}$ &  \multicolumn{2}{c|}{Coordinates (J2000)} & $\sigma_{\mathrm{position}}$  &   L$_{\nu} \times 10^{28}$  \\
\cline{2-3}										     
\cline{6-7}                                                                              
~        &           RA     &      DEC       & (mas) &  (erg s$^{-1}$ Hz$^{-1}$)        &           RA     &      DEC       & (mas) &   (erg s$^{-1}$ Hz$^{-1}$)   \\
\hline						         										       
L1             & 23 39 01.2687    & 36 21 08.5230  & 1.3   &   1.1                      & 07 27 37.6154    & -02 54 54.2540 & 1.4   &     8.4                      \\
L2             & 23 39 01.2674    & 36 21 08.4840  & 1.4   &   1.1                      & 07 27 37.6129    & -02 54 54.2510 & 0.7   &    16.0                      \\
L3             & 23 39 01.2653    & 36 21 08.5260  & 1.2   &   1.2                      & 07 27 37.6113    & -02 54 54.2510 & 1.4   &     8.5                      \\
L4             & 23 39 01.2643    & 36 21 08.5690  & 1.0   &   1.4                      &  ~               & ~              &  ~    &     ~                        \\
L5             & 23 39 01.2624    & 36 21 08.5520  & 0.8   &   1.8                      &  ~               & ~              &  ~    &     ~                        \\
L6             & 23 39 01.2606    & 36 21 08.5720  & 0.7   &   1.9                      &  ~               & ~              &  ~    &     ~                        \\
L7             & 23 39 01.2582    & 36 21 08.5860  & 1.2   &   1.2                      &  ~               & ~              &  ~    &     ~                        \\
C1             & 23 39 01.2665    & 36 21 08.5830  & 0.8   &   0.7                     & 07 27 37.6144    & -02 54 54.2516 & 0.5   &    2.1                       \\
C2             & 23 39 01.2656    & 36 21 08.5895  & 1.1   &   0.5                     & 07 27 37.6129    & -02 54 54.2560 & 0.3   &    3.2                       \\
C3             & 23 39 01.2605    & 36 21 08.5895  & 1.1   &   0.5                     & 07 27 37.6125    & -02 54 54.2540 & 0.3   &    3.1                       \\
C4             & 23 39 01.2599    & 36 21 08.5855  & 1.1   &   0.5                     & 07 27 37.6120    & -02 54 54.2588 & 0.4   &    2.3                       \\
\hline											                      
\end{tabular}
}
\end{center}
\caption{\footnotesize{Parameters estimated from the EVN-observations  \label{ta:param}}}
\end{table}

\vspace{2.5cm}

\end{document}